\newcommand{\bee}{\begin{equation}}
\newcommand{\ene}{\end{equation}}
\newcommand{\beea}{\begin{eqnarray}}
\newcommand{\enea}{\end{eqnarray}}

\documentclass[aps,prl,showpacs,superscriptaddress]{revtex4}
\usepackage{graphicx}
\usepackage{dcolumn}
\usepackage{bm}
\usepackage{amsmath}
\begin{document}
\title{Reply to "Comment on “Novel attractive forces between ions in quantum plasmas"}
\author{P. K. Shukla}
\affiliation{International Centre for Advanced Studies in Physical Sciences \& Institute for Theoretical Physics,
Faculty of Physics \& Astronomy, Ruhr University Bochum, D-44780 Bochum, Germany}
\affiliation{Department of Mechanical and Aerospace Engineering \& Center for Energy Research,
University of California San Diego, La Jolla, CA 92093}
\author{B. Eliasson}
\affiliation{International Centre for Advanced Studies in Physical Sciences \& Institute for Theoretical Physics,
Faculty of Physics \& Astronomy, Ruhr University Bochum, D-44780 Bochum, Germany}
\author{M. Akbari-Moghanjoughi}
\affiliation{Department of Physics, Azarbaijan Shahid Madani University, 51745-406 Tabriz, Iran}

\maketitle
We are responding to the comment of Tyshetskiy and Vladimirov \cite{TV} (hereafter referred to as TV) on a recently published paper by
Shukla and Eliasson \cite{SE12a} (hereafter referred to as SE), with the corresponding Errata \cite{SE12b}, regarding the discovery of a novel attractive force on stationary test ion charges of the same sign shielded by degenerate electrons in a dense quantum plasma. For their purposes, Shukla and Eliasson \cite{SE12a} used the lineraized quantum hydrodynamic (LQHD) equations composed of the electron continuity equation, the inertialess non-relativistic electron momentum equation with electrostatic and quantum forces [e.g. due to the quantum statistical pressure that accounts for electron degeneracy, the quantum recoil effect associated with electron tunneling through the Bohm potential via overlapping of quantum electron wave functions and weak-localization of degenerate electrons, and electron-exchange and electron correlation (EXC and EC) effects due to electron its spin together with the Poisson's equations. In Fourier space the latter yielded the SE potential distribution (Eq. 2 in TV) due to the dielectric constant that resulted by the quantum forces that we described above. TV in their recent evaluation have come up with a misnomer that the short-range (of atomic scales) SE potential is insignificant effect compared to the long-range (in comparison with the Fermi-Thomas length $\lambda_{FT} = v_{Fe}/(\sqrt{3} \omega_{pe})$, where $v_{Fe} =\hbar k_F/m_e$ is the nonrelativistic Fermi electron speed, $\hbar$, the Planck constant divided by $2\pi$, $k_F=(3\pi^2 n_0)^{1/3}$ the Fermi wave-number, $n_0$ the unperturbed electron number-density, $\omega_{pe} =(4\pi n_0 e^2/m_e)^(1/2)$ the electron plasma frequency, $e$ the magnitude of the electron charge, and $m_e$ the rest mass of electrons) Friedel oscillations (FOs), predicted by the Lindhard's dielectric constant $\epsilon (0,k)$ \cite{Lh} in the zero-frequency and zero-Fermi electron temperature limits, by ignoring electron-exchange and electron correlation effects, and which appears as Eq. (4) in TV, where $k$ is a non-normalized wave number. The FOs are caused by discontinuity in the Lindhard'd dielectric function at the Fermi-surface, where $k=2k_F$. In their Fig. 1, TV erroneously compare the SE potential profile \cite{SE12a} deduced from their Eq. (2), depicted by a dashed line, which includes electron-exchange and electron correlation effects, with their oscillatory potential profile (depicted by the solid line) that is obtained by the numerical integration of their Eq. (3) which uses Eq. (4) that completely neglects EXC and EC effects. Such a comparison in Fig. 1 of TV is meaningless because the SE potential distribution (SEPD) profile, as depicted in Ref. \cite{SE12b} significantly differs from the dashed profile in Fig. 1 of TV, besides the fact that the SEPD profile encompasses the physics of the EXC and EC effects that are not included in the Lindhardt's dielectric constant (Eq. 4) which has been used In Eq. (3) of TV.

In the following, we present a corrected analysis that TV gave in their earlier comment \cite{TV}, and demonstrate that the SE potential distribution without the EXC and EC effects (referred to as the reduced SEPD) is a different phenomenon which has nothing to do with the FOs. It should be stressed that both the generalized SEPD and the present reduced SEPD distributions do not assume $k\lambda_{F} \ll 1$ or $k=2k_F$. For our purposes, we shall use the LQHD dielectric constant from Ref. \cite{SE12a} without the EXC and EC effects, which is of the form

\begin{equation}\label{KS}
D^{QHD}  (0, k) \approx 1+\frac{\omega_{pe}^2}{k^2 v_{Fe}^2/3+\hbar^2 k^4/4 m_e^2},
\end{equation}
and the dielectric constant (in the zero-frequency limit) of Lindhard \cite{Lh}, which uses the random-phase-approximation (RPA) \cite{KS,BP}
for noninteracting completely degenerate electron fluids, and which reads \cite{LP}

\begin{equation}\label{RPA}
D^{RPA}(0,k)=1+\frac{3\omega_{pe}^2}{2 k^2 v_{Fe}^2}[1-g(\omega_+)+g(\omega_-)],\hspace{3mm} g(\omega_\pm)
=\frac{m_e(\omega_\pm^2-k^2 v_{Fe}^2)}{2\hbar k^3 v_{Fe}}\ln\left(\frac{\omega_\pm+kv_{Fe}}{\omega_\pm-kv_{Fe}}\right).
\end{equation}
where $\omega_\pm= \pm\hbar k^2/2m_e$. It is interesting to note that Eq. (1) is exactly the same as the one obtained from the quantum kinetic theory's dielectric constant \cite{TT} [viz. the reduced Eq. (14)] in the zero-frequency and zero Fermi electron temperature limits, without any approximation on $k \lambda_F$. Consequently, the electrostatic potential around a stationary test ion charge, in dimensional from, reads as

\begin{equation}\label{phidim}
\varphi (r) = \frac{{4\pi Q}}{{{{\left( {2\pi } \right)}^3}r}}{\mathop{\Im}\nolimits} \left[ {\int_0^\infty  {\frac{{\exp (ikr){d^3}k}}{{{k^2}D(0,k)}}} } \right] = \frac{{2Q}}{{\pi {\lambda _F}r}}\int_0^\infty  {\frac{{\sin (kr)dk}}{{kD(0,k)}}},
\end{equation}
where $\Im$ stands for the imaginary part of integral. It should be noted that the dielectric function can be either $D^{QHD}(0,k)$ or $D^{RPA}(0,k)$ given by Eqs. (\ref{KS}) and (\ref{RPA}), corresponding to the SE or TV schemes, respectively.

It is well known \cite{LP} that the Lindhard dielectric constant, given by Eq. (2), gives rise to the Kohn anomaly \cite{Kohn} due to a singularity at the specific wavelength $\lambda_K =\pi/ k_{F}$ at the electron Fermi surface. Such a singularity is known to produce the periodic local density pattern of the form $n(r) \simeq \sin(2k_{F}r)/r^d$ at distance $r$ around a stationary test charge \cite{SG} far from the charge itself. This is referred to as FOs in  $d$-dimensional compact Fermi electron-sea. In their coment, TV \cite{TV} have presented their numerical simulation results based on the RPA dielectric function (Eq. \ref{RPA}), claiming that the amplitude of SE potential distribution (shown by the solid line in Fig. 1 of TV) is comparable or even less than that of FOs. Our objective here is to point out an obvious mismatch in TV's calculation and inappropriate comparison between the reduced SEPD profile and the potential profile deduced from their Eq. (3) by using Eqs. (4) and (5) which lack essential interaction electron picture. In order to evaluate the outcome of the two approaches on equal footings and to be consistent with TV's scaling, we use the same normalization as in Ref. \cite{TV}. Thus, $k^{-1}$ and $r$ are in units of $\lambda_{F}$. Accordingly, one can cast  Eqs. (\ref{KS}) and (\ref{RPA}) for both the reduced SE (denoted by QHD) and RPA model which ignore the EXC and EC effects. We have

\begin{equation}
D^{QHD}(0,k)=1+\frac{1}{K^2 +\alpha K^4},\hspace{3mm}D^{RPA}(0,k)=1+\frac{1}{2 K^2}\bigg[1+\frac{\sqrt{3}}{2\sqrt{\alpha}K}\left(\frac{\alpha K^2}{3}-1\right)\ln\left|\frac{\sqrt{\alpha} K-\sqrt{3}}{\sqrt{\alpha} K+\sqrt{3}}\right|\bigg],
\end{equation}
where $\alpha=(3\hbar\omega_{pe}/2m_e v_{Fe}^2)^2=(9/\pi^5)^{1/3}(r_0/a_B)=(9/\pi^5)^{1/3}r_s$ ($r_s$ is the Brueckner parameter), for which $r_0\simeq n_0^{-1/3}$ and $a_B=\hbar^2 /e^2 m_e$ are the inter-electron spacing and the Bohr radius of an hydrogen atom, respectively. We note that $\alpha$ here is different from the $\alpha$ in Ref. \cite{SE12a} because of the absence of the electron exchange and electron-correlation potential ($V_{xc}$) in the present analysis.

The normalize potential, $\phi(R)$, defined based on Eq. (\ref{phidim}) in the TV-compatible scaling, may be written as

\begin{equation}\label{phinrm}
\phi (R) = \frac{2}{{\pi R}}\int_0^\infty  {\frac{{\sin (KR)}}{{KD(0,K)}}} dK,
\end{equation}
where the electrostatic potential $\phi (R)$ is normalized by $Q/\lambda_{F}$, with $Q$ being the ion charge state, and $D(0,k)$ the static dielectric constant for the reduced SE and RPA cases.

\begin{figure}[htb]
\label{Figure 1}
\includegraphics[width=10cm]{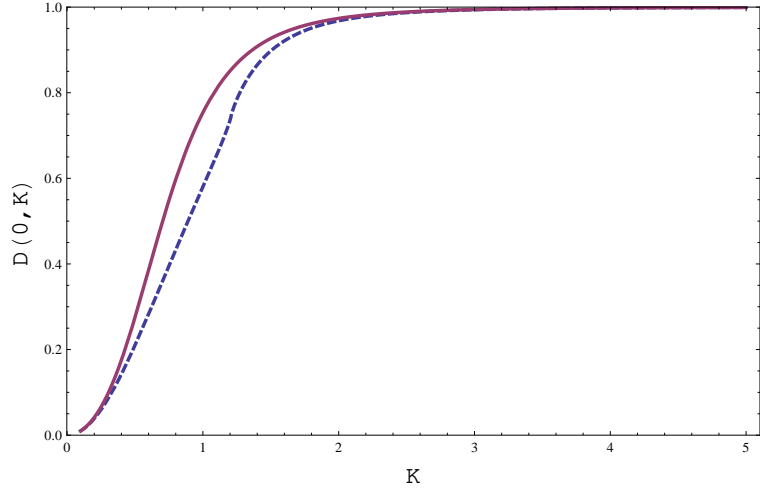}
\caption{(Color online) The inverse of the dielectric constant versus the parameter $K$ using the reduced QHD (solid) and RPA (dashed) cases for $\alpha=2.06$ ($n_0=9.63\times 10^{22}$cm$^{-3}$) corresponding to Fig. 1 of TV.}
\end{figure}

\begin{figure}[htb]
\label{Figure 2}
\includegraphics[width=17cm]{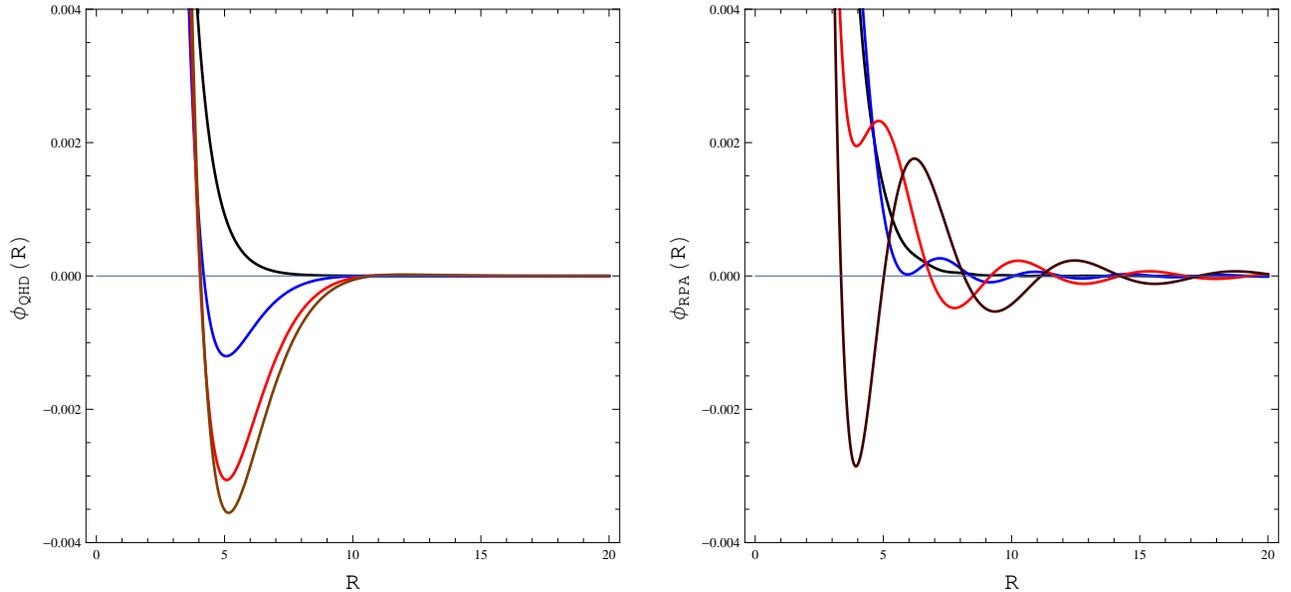}
\caption{(Color online) The profile of the potentials as a function of $R$ for different values of the electron number density $n_0$ for
the reduced SE (QHD model) (left) and RPA (right) models for different electron number-densities $n_0=(7.94, 9.63^*, 21.3, 104) \times 10^{22}$ cm$^{-3}$ (the star mark denotes the number-density used in TV's comment) shown by (Brown, Red, Blue, Black) colors, respectively. Similar colors in the two models correspond to the same electron number density.}
\end{figure}

Figure 1 depicts the inverse of the dielectric constant, $1/D(0,k)$, for the reduced QHD and RPA cases for a given electron number density $n_0=9.63\times 10^{22}$ cm$^{-3}$, which corresponds to $\alpha\simeq2.06$ ($r_s\simeq 6.68$) used in Fig. 1 of TV's comment \cite{TV}. It is noted that in the long-wavelength limit $1/D(0,\infty)=1$ is fulfilled for both cases, as expected. The Kohn-anomaly due to the singularity in the Fermi electron-liquid dispersion embedded in Eq. (2), which leads to weak density oscillations (FOs) around an impurity test charge, occurs exactly in the middle region of the dashed curve (the RPA case) at $k=2k_{Fe}$, where it seems to be broken.

Figure 2, on the other hand, exhibits the variation of the potential profiles for the reduced QHD (the left plot) and RPA (the right plot) cases for different electron number densities of our concern in this reply. These profiles from thick to thin curves (the highest to the lowest electron number densities) correspond to $\alpha\simeq\{0.19,0.93,2.06,2.5\}$ or $r_s\simeq\{0.62, 3, 6.68, 8.1\}$, respectively. The potential distribution profile corresponding to the electron number density of $n_0=9.93\times 10^{22}$ cm$^{-3}$ (denoted by an asterisk symbol in the plots) is the one given by TV. Indeed, the expected oscillations from the RPA theory are evident from Fig. 2. However, while there appears to be a pattern for the electric potential depth and its location in the reduced QHD case, in the RPA case no such pattern is found at least for characterizing the first minimum in the oscillations. Such lack of consistent pattern can be marked as a disadvantage for a theory. While TV fail to simulate the reduced QHD potential minimum location correctly in their paper, they also underestimate its depth as is apparent when comparing Fig. 2 for $n_0=9.93\times 10^{22}$ cm$^{-3}$ case here with that of Fig. 1 in TV's comment paper \cite{TV}. The first positive local minima of the corresponding, $\alpha=2.06$, the potential profile for the RPA case is not even shown in Ref. \cite{TV}. The location of this positive minimum (not shown in Ref. \cite{TV}) is around the minimum of the corrected potential profile for the reduced QHD case (e.g. see Fig. 2 here) for the same electron number density. It is found that
for all electron number density values used here (including the one used Ref. \cite{TV}) the amplitudes of the RPA profiles are lower than that of the reduced QHD case, which is in contrast to the findings of Ref. \cite{TV}. Only a somewhat qualitative agreement between the  reduced QHD and RPA cases seems to appear in Fig. 2 for the lowest metallic density of $n_0=7.94\times 10^{22}$ cm$^{-3}$ and the highest electron number density $n_0=104\times 10^{22}$ cm$^{-3}$, with the Thomas-Fermi-like charge screening.

On the other hand, it has to be mentioned that the results of the RPA case without considering the important electron interaction features, such as EXC and EC effects, are highly unreliable. TV claim that such an extension of the RPA theory is being elaborated by them. However, such an extension has in fact been already carried out by Singwi {\it et.al.} \cite{SW1,SW2}, who have proven that the Lindhard dielectric constant \cite{Lh,LP}, obtained by using  the RPA, is only adequate for the description of the plasmon excitation modes in the long-wavelength limit and its applicability is limited to the high electron number density. This theory has also been shown to lead to nonphysical negative pair-distribution function for the entire range of the metallic density, which is caused by the lack of the local field correction and short-range effects in the RPA theory \cite{SW1}. The screening density (which is proportional to the screening electrostatic potential) for the RPA and improved Hubbard and Singwi models has been shown for $r_s=3,6$ values in Figs. 6 and 7 of Ref. \cite{SW2}. At a first glance of these Figs, it appears that the improved screening tends to deepen the first valley making it look more like those of the reduced QHD rather than those of the RPA, as shown in Fig. 2. Let us closely inspect Figs. 6 and 7 of Ref. \cite{SW2} against the potential profiles displayed  in Fig. 2 for the electron number densities $n_0=9.63\times 10^{22}$ cm$^{-3}$ and $n_0=21.3\times 10^{22}$ cm$^{-3}$ corresponding to Brueckner parameters of $r_s=3$ and $r_s=6.68$, respectively. It is observed that in improved models while the location of the principal minimum is almost unchanged, its depth increases with a decrease in the electron number
density. These findings are in complete agreement with the reduced QHD findings for the electric  potential profile, as  shown here in Fig. 2. It also sharply contrasts the findings of the RPA result that is presented in our Fig. 2. It is evident that the noninteracting RPA model used in Ref. \cite{TV} does not lead to physically consistent results. For metallic densities of $n_0=9.63\times 10^{22}$ cm$^{-3}$ and $n_0=21.3\times 10^{22}$ cm$^{-3}$ it does not even present a well pronounced principal negative potential valley, a characteristics of the quantum plasma screening.
Such a well pronounced negative potential valley can even be found in density functional theory (DFT) simulation results \cite{CH} of quantum plasma systems, which deepens the potential well by including  local density corrections.

In summary, we conclude that only a quantum kinetic theory which includes all quantum forces, as described in Ref. \cite{SE12a}, must be used
to obtain physically consistent results for the quantum screening. The RPA model used by TV is not consistent with the LQHD, DFT and improved Hubbard and Singwi models and should therefore be discarded. There are oscillations in the screened potential tail of a quantum electron fluid
due to the Kohn-anomaly, insignificant compared to the well-pronounced reduced SE effect. Such insignificant oscillations can also be probed
by extra zooming-in of the reduced SE potential's tail, which might appropriately account for the FOs phenomenon.


\begin{thebibliography}{}
\bibitem{TV} Yu. Tyshetskiy and S. V. Vladimirov, Comment on "Novel Attractive Force between Ions in Quantum Plasmas",
submitted to Phys. Rev. Lett. (under review); see also E-print: arXiv:1212.4286v1 [physics.plasm-ph] 19 Dec 2012.
\bibitem{SE12a} P. K. Shukla and B. Eliasson, Phys. Rev. Lett. {\bf 108}, 165007 (2012).
\bibitem{SE12b} P. K. Shukla and B. Eliasson, Phys. Rev. Lett.  {\bf 108}, 219902 (E) (2012); {\it ibid.} {\bf 109}, 019901 (E) (2012).
\bibitem{Lh} J.Lindhard, Kg. Danske Videnskab. Selskab, Mat. Fys. Medd. {\bf 28}, No.8, 1 (1954).
\bibitem{KS} Yu. L. Klimontovich and V. P. Silin, Dokl. Akad. Nauk SSSR {\bf 82}, 361 (1952); Zh. Eksp. Teor. Fiz. {\bf 23}, 151 (1952).
\bibitem{BP} D. Bohm and D. Pines, Phys. Rev. {\bf 92}, 609 (1953).
\bibitem{LP} E. M. Lifshitz and L. P. Pitaevskii, {\em Physical Kinetics} (Butterworth--Heinemann, Oxford, 1981).
\bibitem{TT} N. L. Tsintsadze and L. N. Tsintsadze, Europhys. Lett. {\bf 88}, 35001 (2009).
\bibitem{Kohn} W. Kohn, Phys. Rev. Lett {\bf 2}, 393 (1959).
\bibitem{SG} George E. Simion and Gabriele F. Giuliani, Phys. Rev. B {\bf 72}, 045127 (2005).
\bibitem{SW1} K. S. Singwi, M. P. Tosi and R. H. Land, Phys. Rev. {\bf 176}, 589 (1968).
\bibitem{SW2} K. S. Singwi, M. P. Tosi and R. H. Land, Phys. Rev. B {\bf 1}, 1044 (1970).
\bibitem{CH} Jeng-Da Chai and John D. Weeks, J. Phys. Chem. B {\bf 108}, 6870 {2004}.
\end{thebibliography}
\end{document}